\documentclass{elsart}
\usepackage{epsfig}
\usepackage{enumerate}

\begin{document}
\begin{frontmatter}

\title{The first shall be last: selection-driven minority becomes majority}

\author{Nuno Crokidakis$^{1,2}$}
\thanks{nuno@if.uff.br}
\author{and Paulo Murilo Castro de Oliveira$^{1,3}$}
\thanks{pmco@if.uff.br}
\address{
$^{1}$Instituto de F\'{\i}sica, Universidade Federal Fluminense, Niter\'oi/RJ, Brazil \\
$^{2}$Departamento de F\'isica, PUC-Rio, Rio de Janeiro/RJ, Brazil \\
$^{3}$National Institute of Science and Technology for Complex Systems, Brazil}

\maketitle

\begin{abstract}
\noindent
Street demonstrations occur across the world. In Rio de Janeiro, June/July 2013, they reach beyond one million people. A wrathful reader of \textit{O Globo}, leading newspaper in the same city, published a letter \cite{OGlobo} where many social questions are stated and answered Yes or No. These million people of street demonstrations share opinion consensus about a similar set of social issues. But they did not reach this consensus within such a huge numbered meetings. Earlier, they have met in diverse small groups where some of them could be convinced to change mind by other few fellows. Suddenly, a macroscopic consensus emerges. Many other big manifestations are widespread all over the world in recent times, and are supposed to remain in the future. The interesting questions are: 1) How a binary-option opinion distributed among some population evolves in time, through local changes occurred within small-group meetings? and 2) Is there some natural selection rule acting upon? Here, we address these questions through an agent-based model.

\end{abstract}
\end{frontmatter}

Keywords: Opinion Dynamics, Collective Phenomenon, Agent-based model, Computer Simulation

\section{Introduction}

\qquad Social dynamics have been studied through statistical physics techniques in the last twenty years and are now part of the new branch of physics called Sociophysics \cite{galam_book,sen_book}. This recent research area uses tools and concepts of the physics of disordered matter and more recently from the network science to describe some aspects of social and political behavior \cite{loreto_rmp}. From the theoretical point of view, opinion models are interesting to physicists because they present correlations, order-disorder transitions, scaling and universality, among other typical features of physical systems \cite{loreto_rmp}.
 
One of the most studied models of opinion formation is the majority-rule model \cite{galam1,galam2,redner}. It considers a population of $N$ agents, each carrying one of two possible opinions $A$ or $B$. A group of $g$ agents (where $g$ is an odd number) is picked at random and all agents in the group adopt the state of the local majority. These rules are repeated until the system reaches a final state of consensus, where all agents are $A$- or $B$-supporters. In this work we introduce a mechanism in the majority-rule model that limits the capacity of persuasion of the majorities, and study the impact of this limitation on the opinion formation.

This work is organized as follows. In Section 2 we present the model, define its microscopic rules and discuss the numerical results. Our conclusions are presented in Section 3.


\section{Model and Results}

\qquad We consider a population of $N=n_{A}+n_{B}$ agents with opinions $A$ or $B$ concerning a given subject. The initial fraction of $A$-supporters is $D=n_{A}/N$ at $t=0$. Each agent $j$ has its own ``convincing power'' (CP), a real number $P_{j}$ initially sorted uniformly inside a given interval $[-C,C]$. A group of $g=3$ agents is randomly chosen \footnote{In this sense, the model has no spatial structure, i.e., all agents have the same probability to be chosen.}. If there is a majority of two, say $i$ and $j$, in favor of one opinion and the average convincing power $\bar{P}=(P_{i}+P_{j})/2$ of this pair is larger than that of the third agent, say $k$, this last flips his opinion and the winning pair $(i,j)$ increases its convincing powers, i.e., we update $P_{i}\to P_{i}+1$ and $P_{j}\to P_{j}+1$. Otherwise, nothing occurs. One time step in the model consists of the application of those rules $N$ times. Notice from the microscopic rules that the agents' capacities of persuasion are limited, compared with the traditional pure majority-rule: only if $\bar{P} > P_{k}$ the majority succeeds in convincing the minority within the tossed $g$-group. In addition, the updating of the CPs models the tendency for a given agent who succeeds in persuading many others to increase his persuasion ability \cite{meu_pla}. This new ingredient introduces a historical long-term memory into the dynamical evolution which turns it non-trivial, with its non-ergodicity exhibited since beginning \footnote{In spite the previous footnote, the model cannot be considered ``mean-field-like". One cannot replace the tossed $g$ agents by any kind of global average.}. Thus, the inclusion of the above rules is a simple form to capture this real-world characteristic.

Notice also that the distribution of CPs (uniform inside $[-C,C]$ at beginning or any other future form) can be freely translated by a constant displacement (as a whole), and thus can include even negative numbers. This is due to the fact that only the relative difference among the convincing powers ($\bar{P}$ and $P_{k}$) matters.

\begin{figure}[t]
\begin{center}
\vspace{0.2cm}
\includegraphics[width=0.4\textwidth,angle=0]{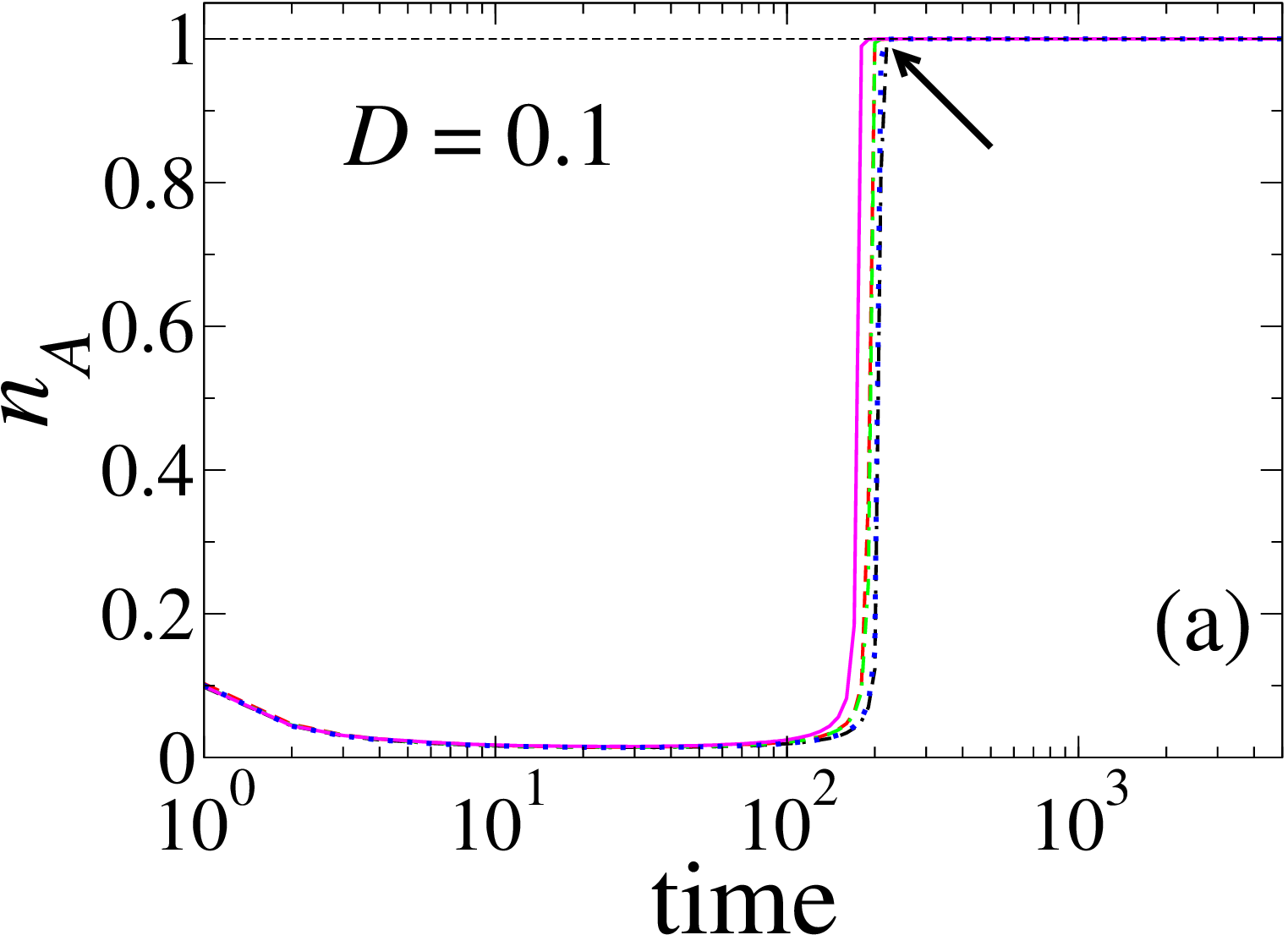}
\hspace{0.4cm}
\includegraphics[width=0.4\textwidth,angle=0]{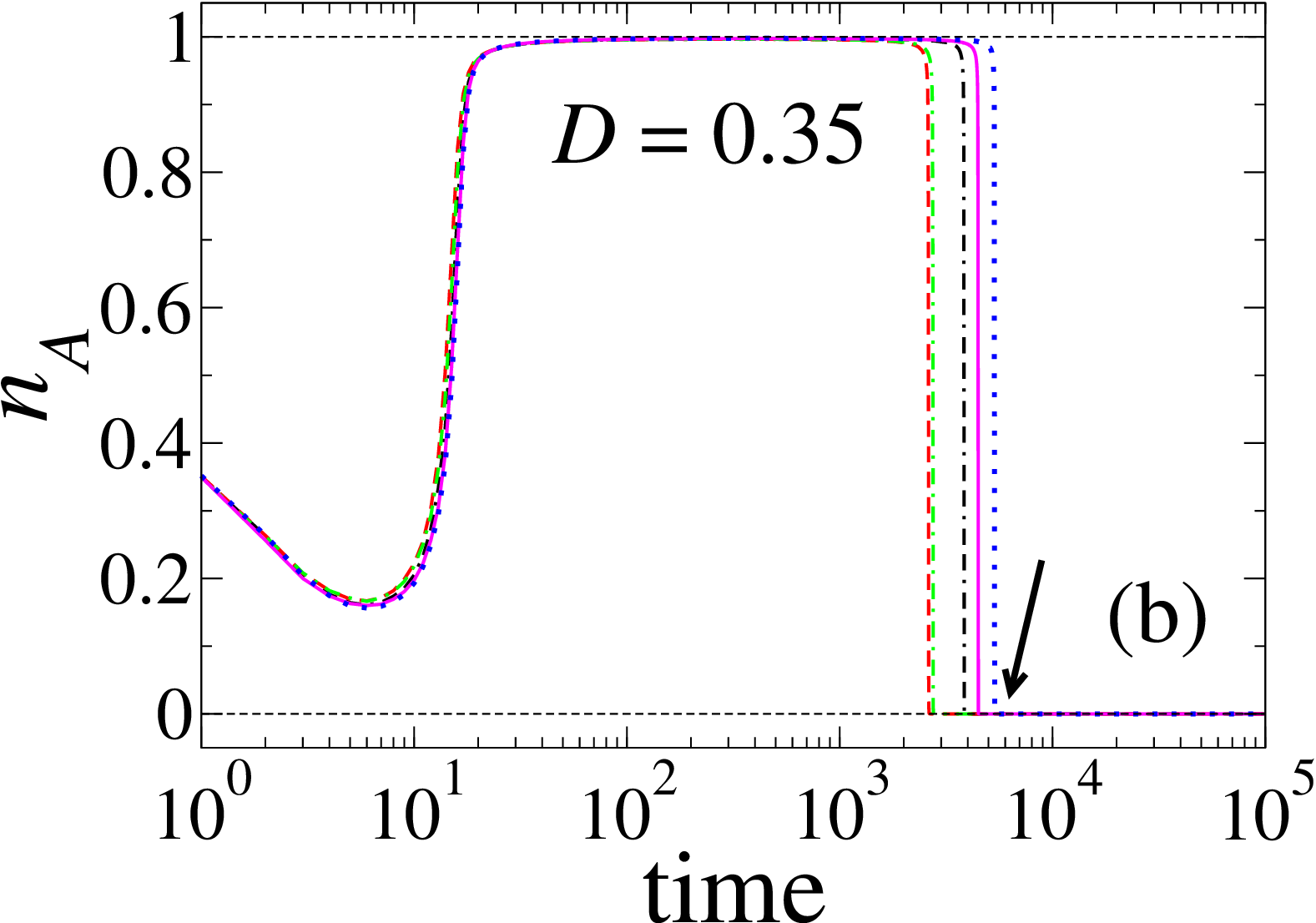}
\\
\vspace{0.65cm}
\includegraphics[width=0.4\textwidth,angle=0]{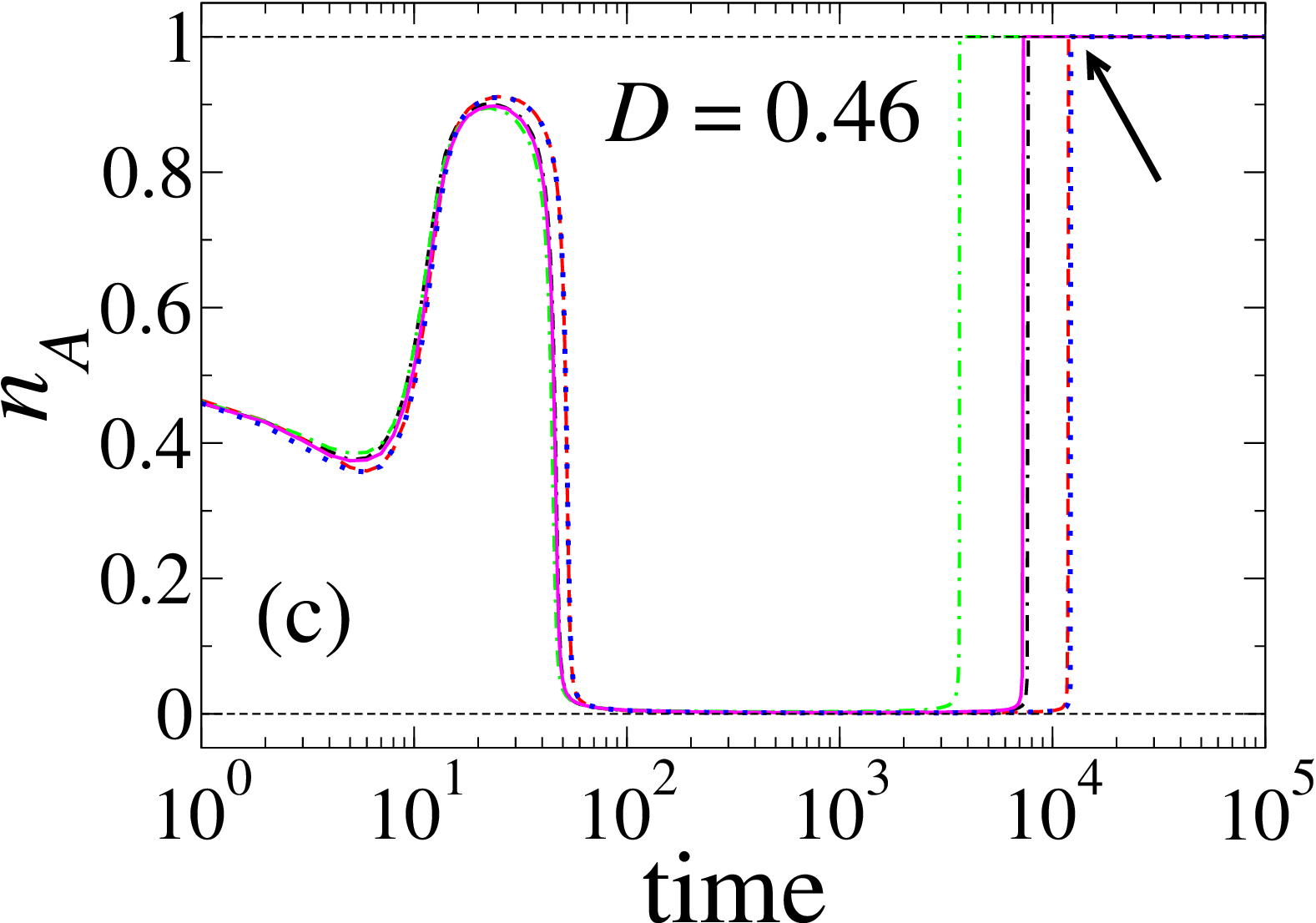}
\end{center}
\caption{(Color online) Time evolution of the fraction $n_{A}$ of $A$-opinions for population size $N=10^{5}$. Results are for different initial fractions of $A$-supporters, namely $D=0.1$ (a), $D=0.35$ (b) and $D=0.46$ (c). Arrows indicate which ($A$ or $B$) final consensus is reached. Each curve represents a single realization of the dynamics.}
\label{fig1}
\end{figure}

In this work we will present results for $C=5$, but the behavior of the model does not depend qualitatively on the value of $C$. One can start studying the time evolution of one of the two fractions of opinions $A$ or $B$, represented by Ising variables $s=+1$ and $s=-1$, respectively. Since the system is symmetric around $D=0.5$, in Fig. \ref{fig1} we exhibit the time evolution of the fraction $n_{A}$ of $A$-opinions for typical values of $D<0.5$ and population size $N=10^{5}$. In opposition to what occurs in the pure majority-rule model \cite{galam1,galam2}, the initial minority can grow, eventually dominating all the population. For the case $D=0.1$, Fig. \ref{fig1} (a), $90\, \%$ of the population starts with opinion $B$ ($s=-1$). After few time steps the number of individuals with the minority opinion $A$ ($s= +1$) decreases, but the majority opinion $B$ does not reach $100\, \%$ of the population. In this case, the system remains in this state near the consensus with $n_{A}=0$ (and $n_{B}=1$) for a long time, but after that the number of individuals with $s=+1$ starts to increase and suddenly grows until the consensus state with $n_{A}=1$. This behavior occurs for all simulated samples (5 for the plot in Fig. \ref{fig1}, thousands in total). In the case $D=0.35$, one can see in Fig. \ref{fig1} (b) that there is an oscillation with respect to the majority opinion: initially the opinion $A$ ($s=+1$) is the minority one, but after some time steps a reversion occurs and the opinion $B$ ($s= -1$) becomes the (second) minority. In this case, the state $s=-1$ is shared by a very small fraction of the population, less than the previous $A$-minority (near $t=10-20$, see Fig. \ref{fig1} (b)). This new global $B$-minority opinion grows and dominates all the population near $t=10^{3} - 10^{4}$. The first $A$-minority is not  sufficiently small to win the final consensus, in opposition to what occurs with the second, smaller $B$-minority. This kind of oscillation and reversion of the majority opinion occurs three times in the case $D=0.46$, as exhibited in Fig. \ref{fig1} (c).

\begin{figure}[t]
\begin{center}
\vspace{0.2cm}
\includegraphics[width=0.3\textwidth,angle=0]{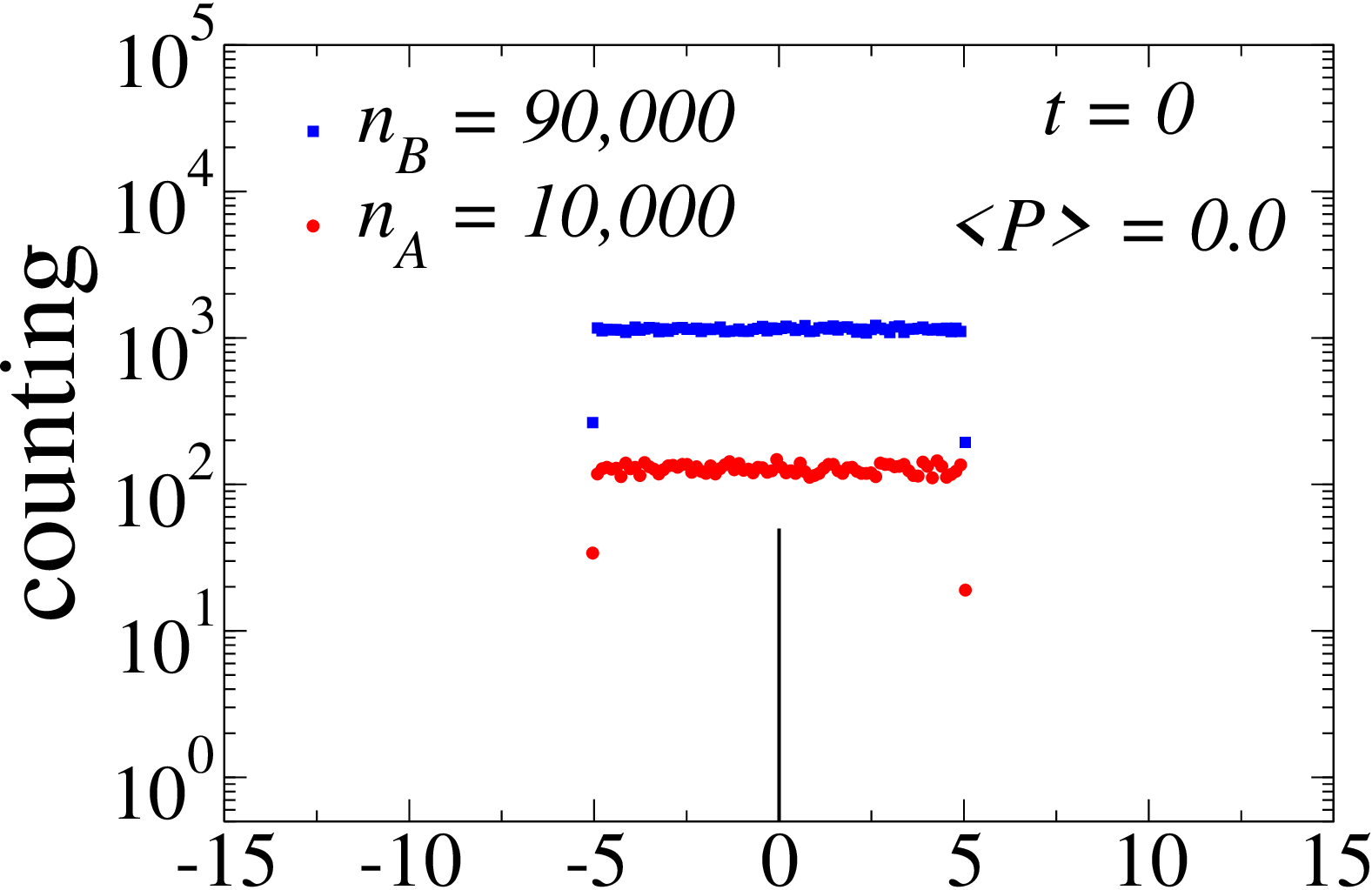}
\hspace{0.2cm}
\includegraphics[width=0.285\textwidth,angle=0]{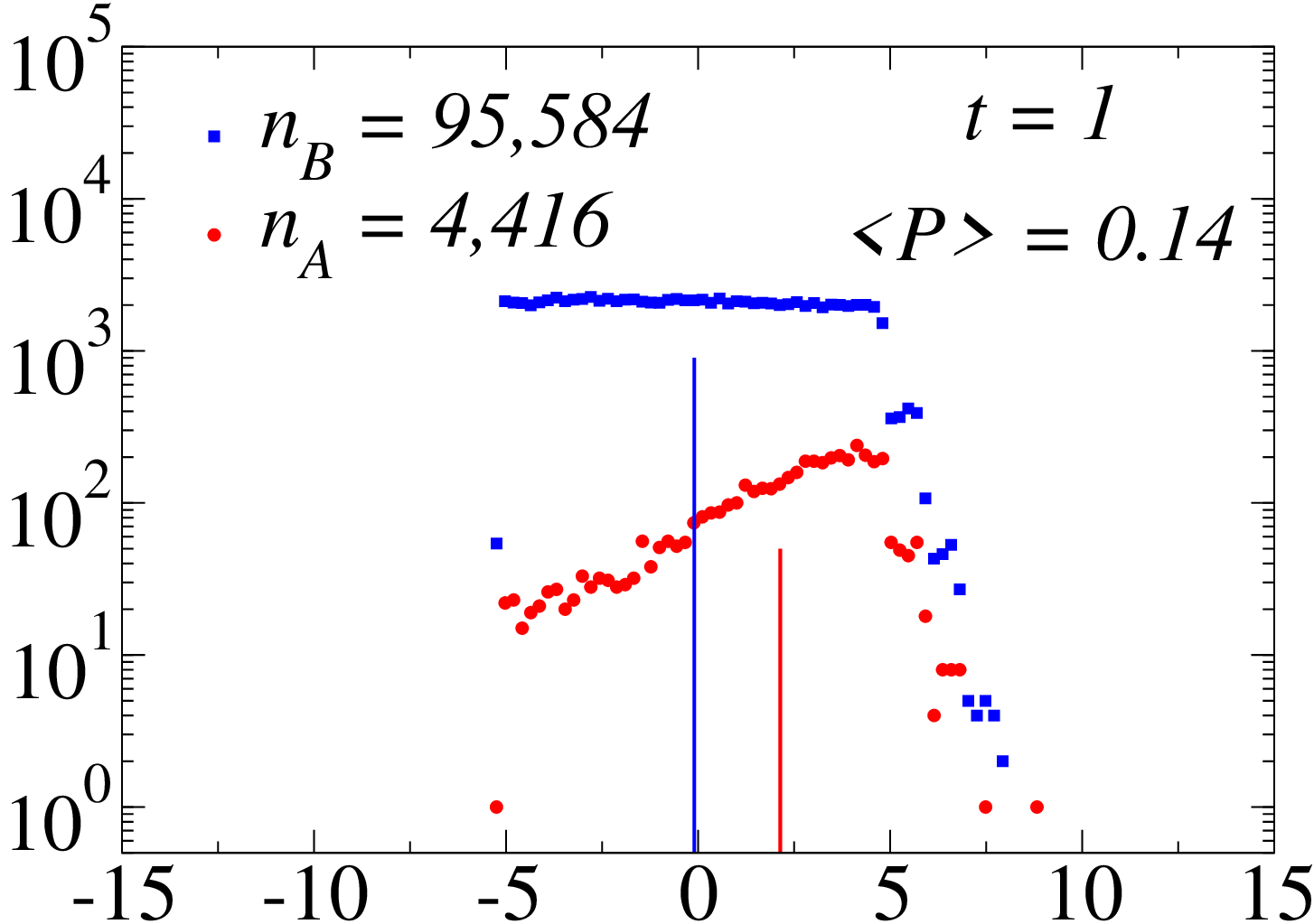}
\\
\vspace{0.39cm}
\includegraphics[width=0.3\textwidth,angle=0]{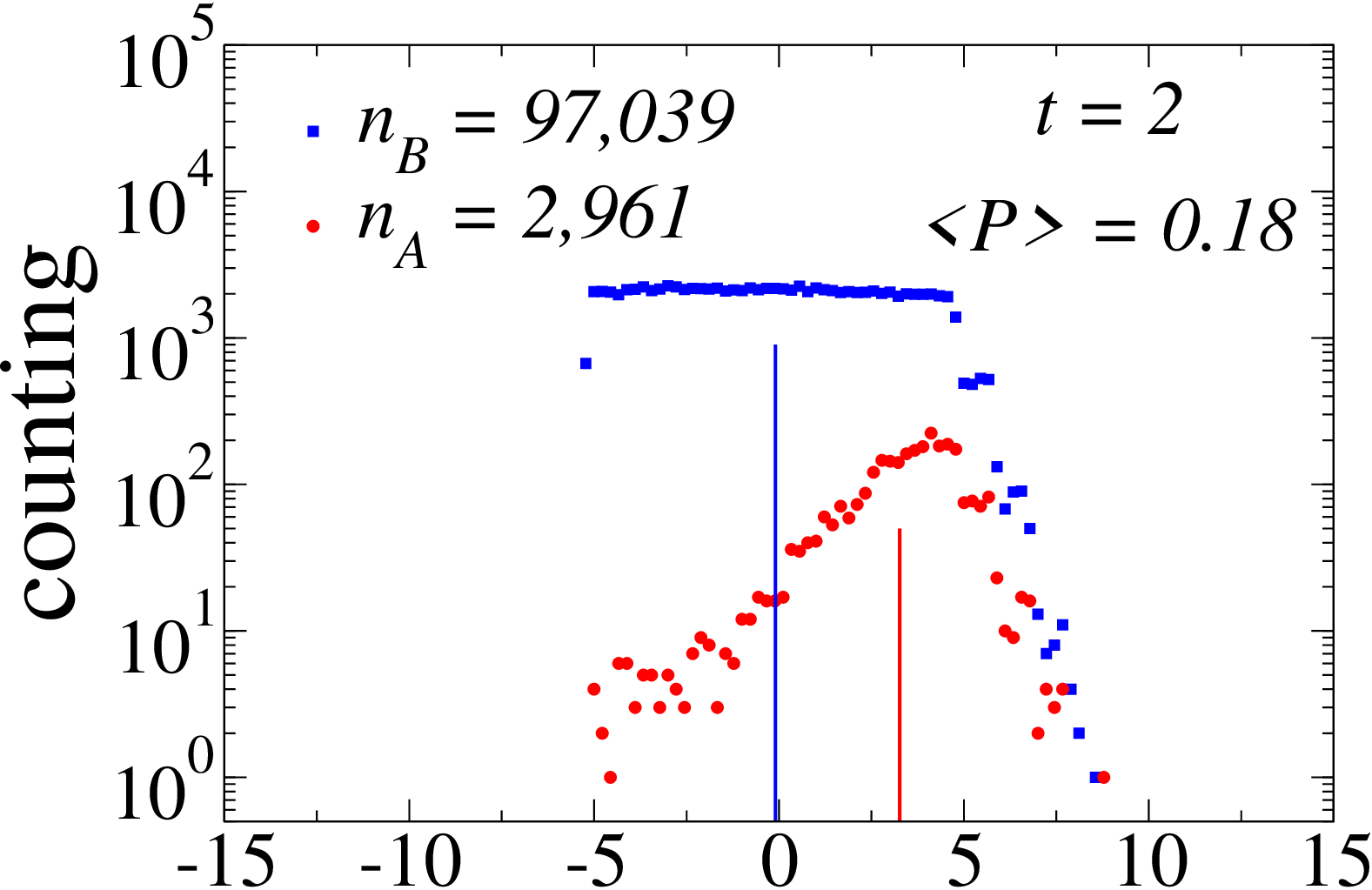}
\hspace{0.2cm}
\includegraphics[width=0.285\textwidth,angle=0]{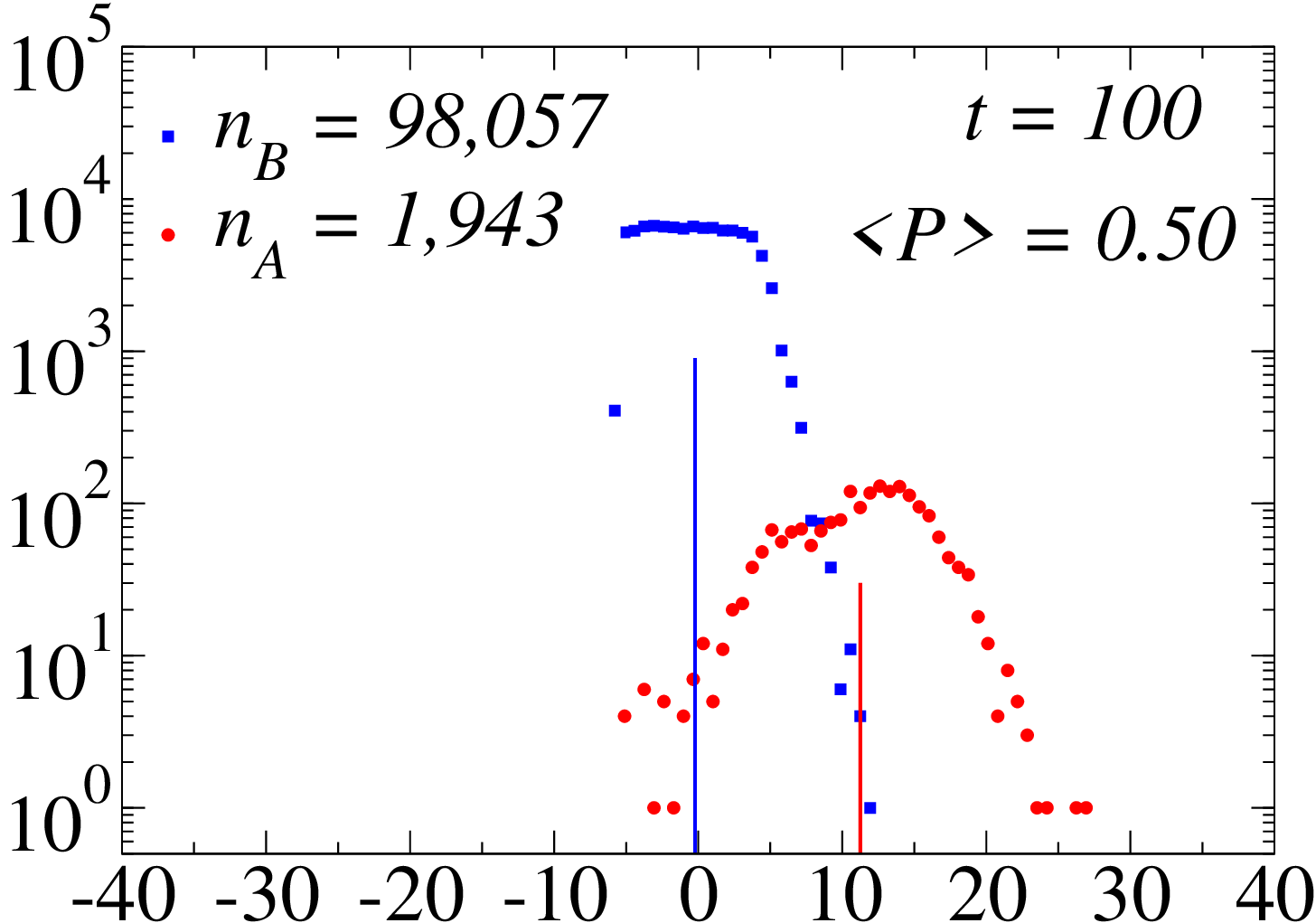}
\\
\vspace{0.39cm}
\includegraphics[width=0.3\textwidth,angle=0]{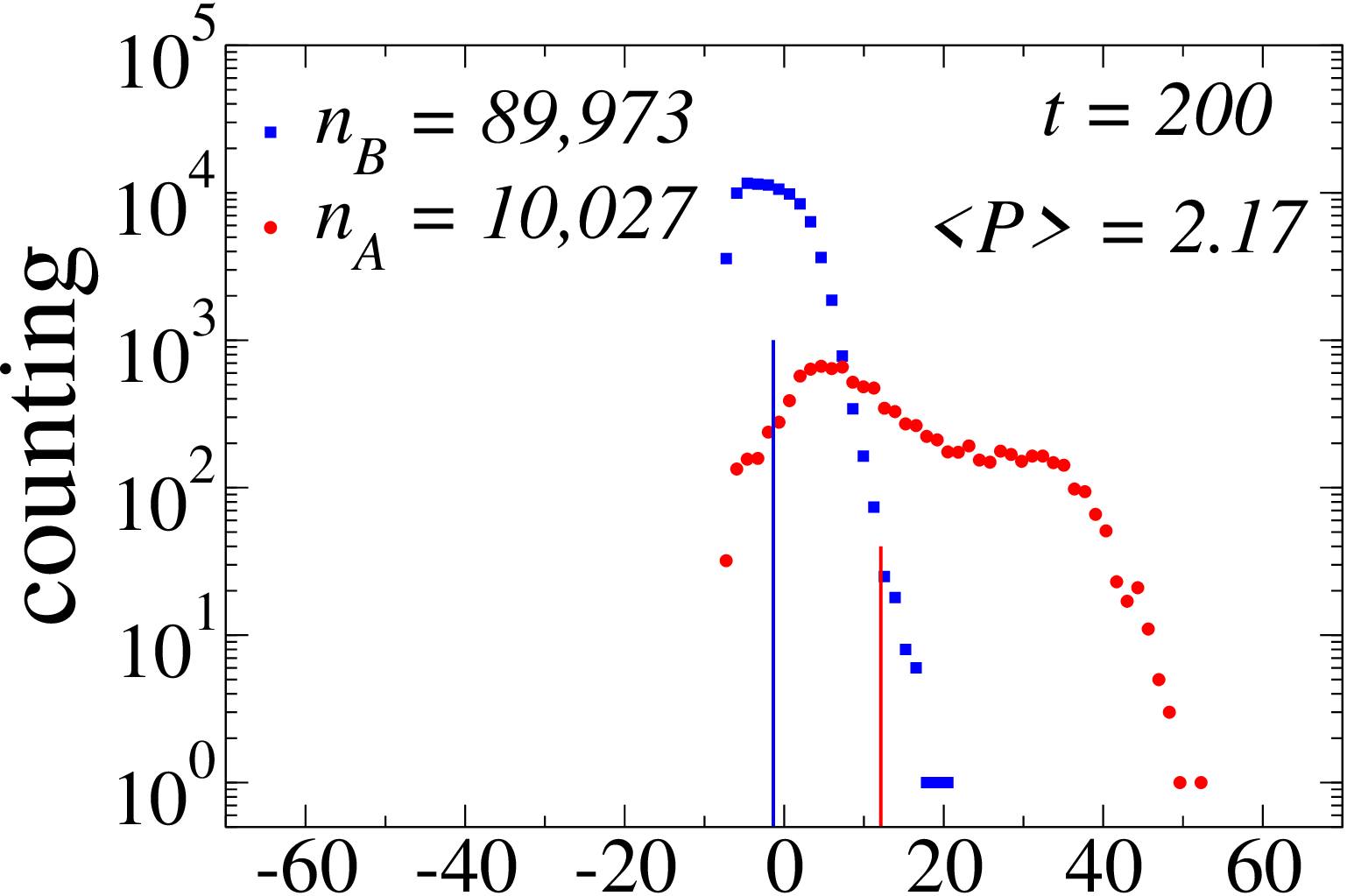}
\hspace{0.2cm}
\includegraphics[width=0.285\textwidth,angle=0]{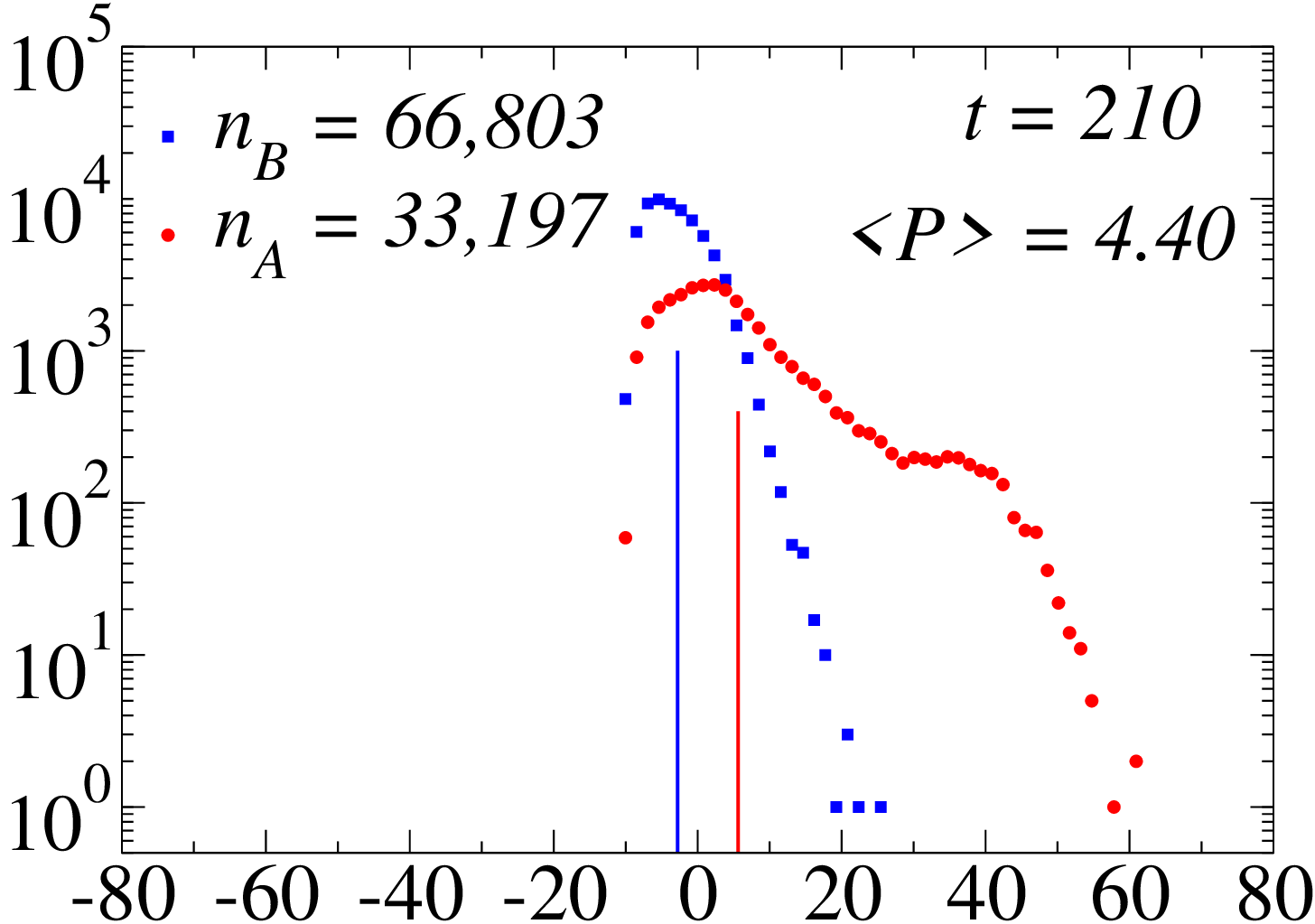}
\\
\vspace{0.45cm}
\includegraphics[width=0.3\textwidth,angle=0]{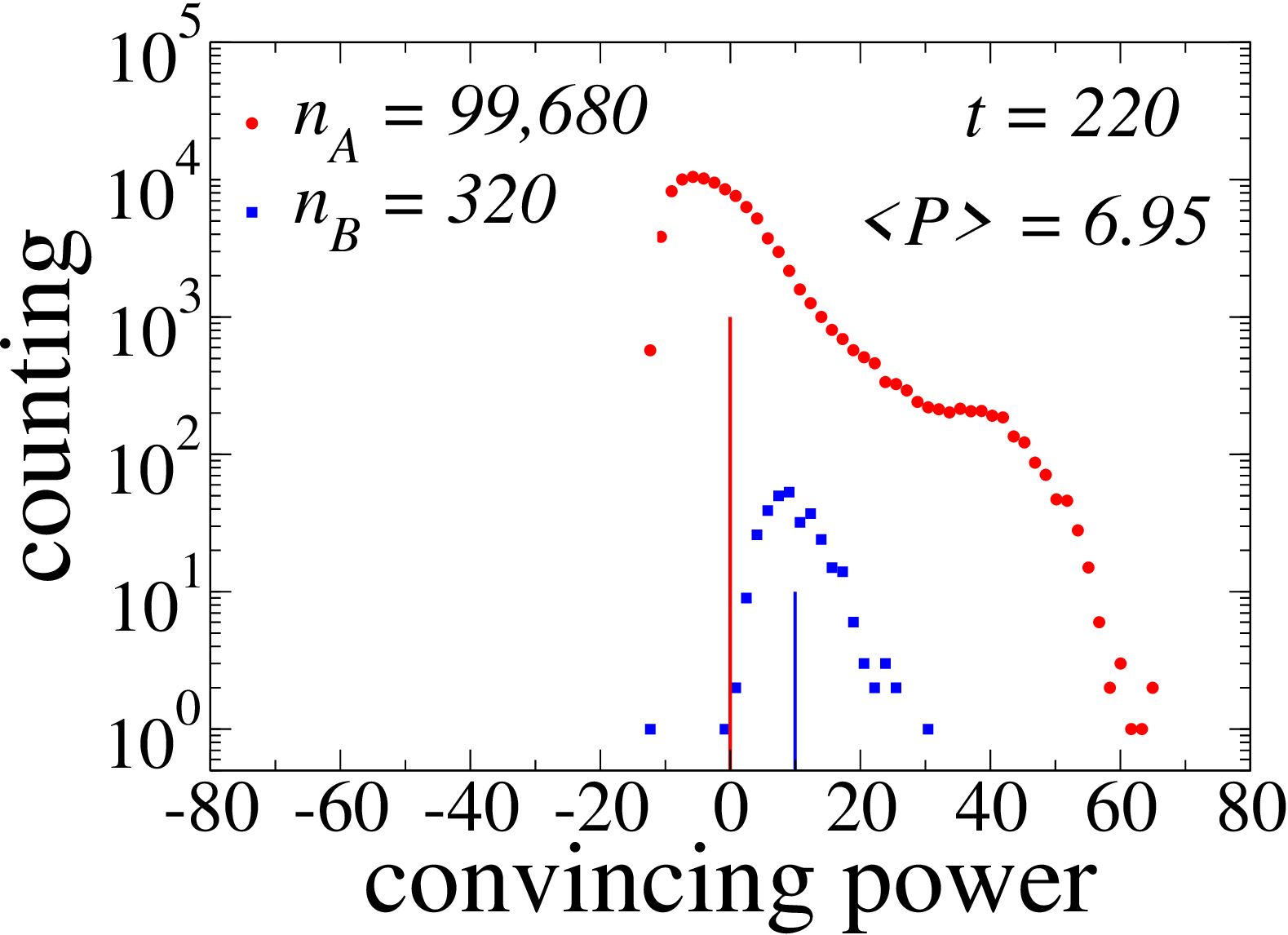}
\hspace{0.2cm}
\includegraphics[width=0.285\textwidth,angle=0]{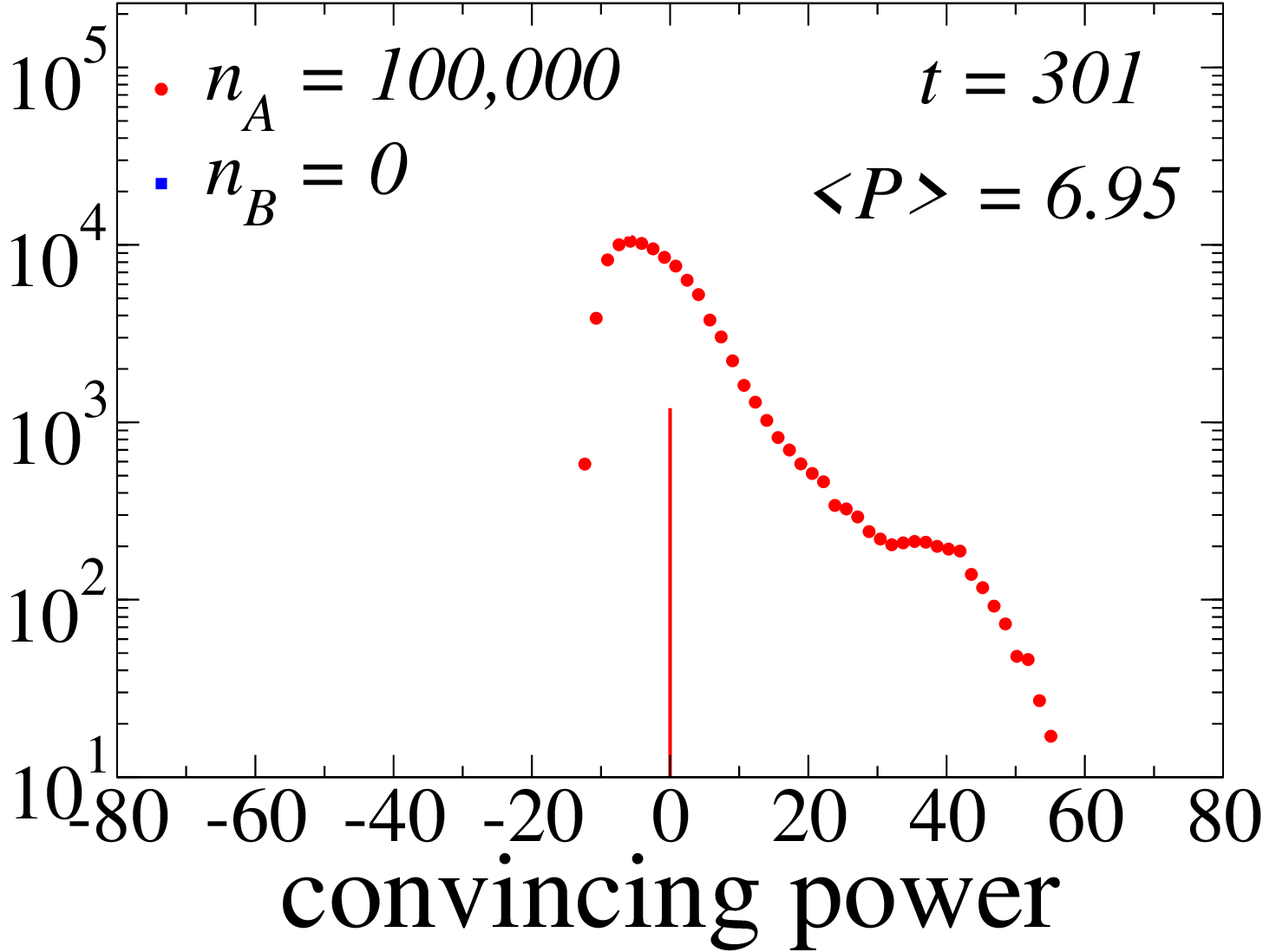}
\end{center}
\caption{(Color online) Time evolution of $A$- and $B$- distributions of the convincing powers for $D=0.1$ and population size $N=10^{5}$, for distinct time steps $t$. The average value $\langle P\rangle$ of the whole distribution is translated to the origin in all plots (its original value is written). $A$- and $B$- averages are indicated by spikes. Observe, already for $t=1$, the sloped behavior of the lower distribution indicating the presence of a strong selection pressure among the minority, absent among the majority, upper distribution. That is why opinion $A$ eventually dominates the population, in spite of being the initial minority.}
\label{fig2}
\end{figure}

The final victory of the (sufficiently small) minority is a consequence of the limited persuasion considered in the majority-rule model, promoting selection among the minority group but not among the majority, a Darwinian bottleneck as explained below. Let's take the case $D=0.1$ for which we have verified that the lower values of the fraction of $A$-opinions are close to $n_{A}=0.02$, i.e., $\approx 98\, \%$ of the population are favorable to opinion $B$ ($s=-1$). However, as agents forming the groups are randomly chosen, we have in this case many groups of $g=3$ individuals with $s=-1$, which does not change their convincing powers. This fact explains why the system is kept for a long time in a quasi-stationary state. During this period, individuals with minority opinion, when rarely chosen to form a majority in a $g$-group, can persuade individuals with $s=-1$, upgrading their convincing powers. A tiny minority opinion ($A$, in this case) group is nucleated with high convincing powers. They cannot be persuaded to change to the majority opinion ($B$), and eventually dominate the whole population.

One can better understand the behavior presented in Fig. \ref{fig1} if we follow the time evolution of the agents' CPs $A$- and $B$-distributions. In Fig. \ref{fig2} we exhibit a typical realization for $D=0.1$ and $N=10^{5}$ agents, at certain time steps $t$. The initial majority is favorable to opinion $B$ (90,000 agents, see Fig. \ref{fig2}). At $t=0$ we have a uniform distribution of CPs with average $\langle P\rangle=0$. After 1 time step one can see that the CP distribution for the current majority ($B$-supporters) remains similar as at $t=0$, no selection. But the CP distribution for the current minority ($A$-supporters) changes: a selection towards high CP is at work, the quoted Darwinian bottleneck. In evolutionary biology, populations suffering a drastic decrement in size tend to have a set of few individuals with high reproductive activities. This leads to a corresponding decrement in genetic diversity, the founder effect. Here, a small group of $A$-supporters (a minority among the minority) consistently increase their convincing powers, surpassing that of all $B$-supporters. This small group can no longer be persuaded, and keeps the minority alive. Finally, an inversion of the majority opinion occurs between $t=210$ and $t=220$, until the consensus with $A$-opinion is reached.

\begin{figure}[t]
\begin{center}
\vspace{0.5cm}
\includegraphics[width=0.3\textwidth,angle=0]{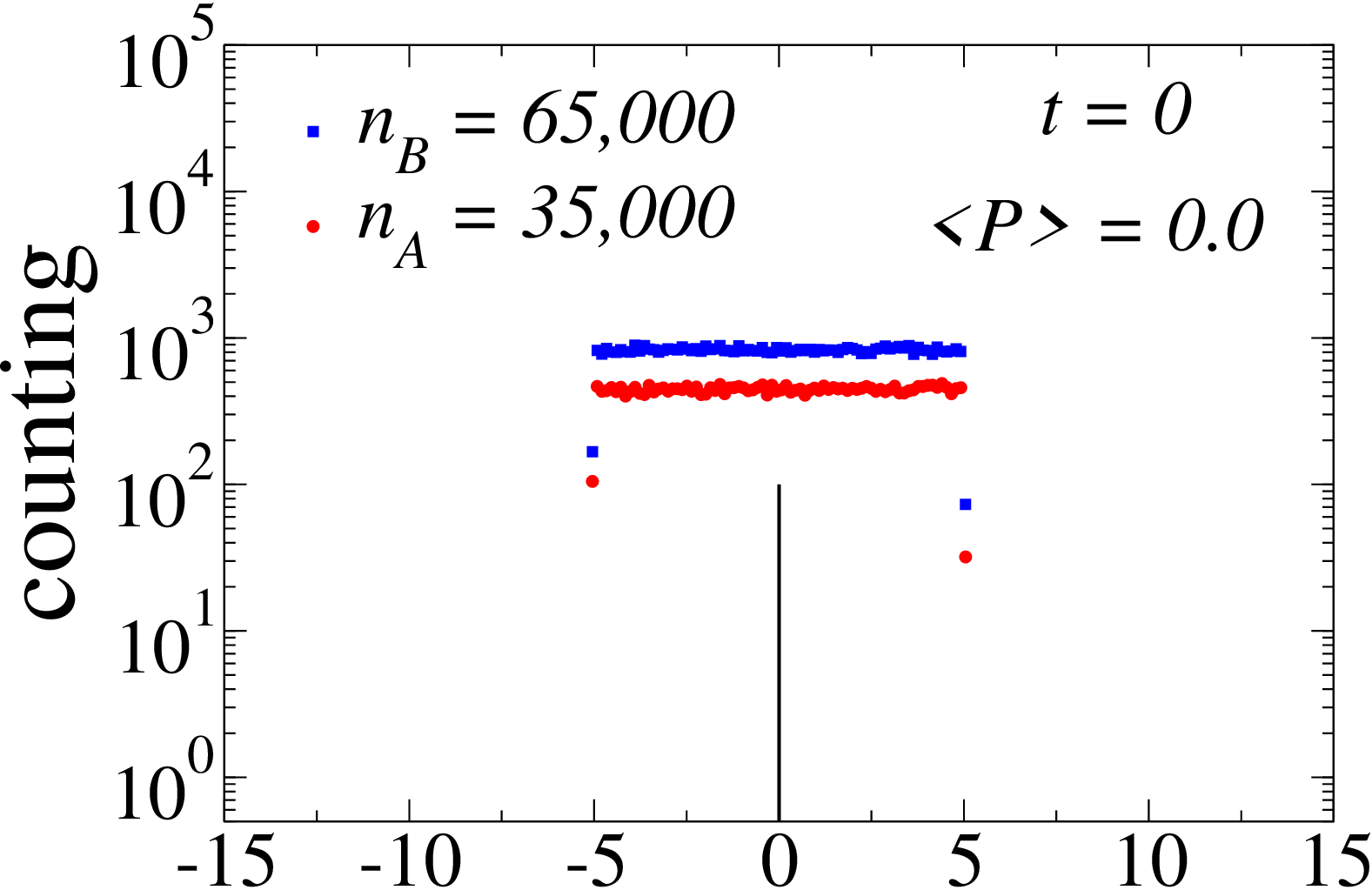}
\hspace{0.2cm}
\includegraphics[width=0.285\textwidth,angle=0]{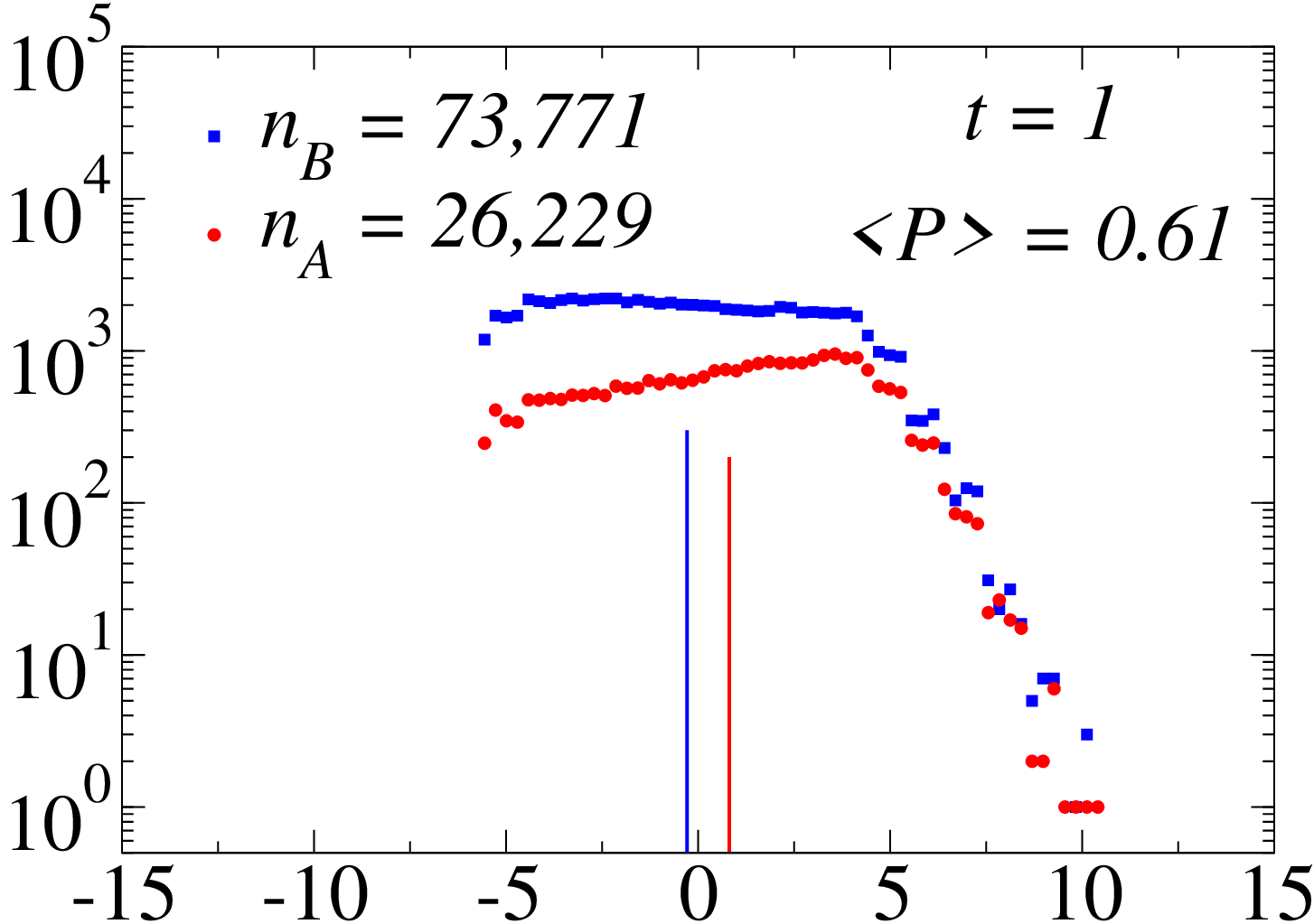}
\\
\vspace{0.48cm}
\includegraphics[width=0.3\textwidth,angle=0]{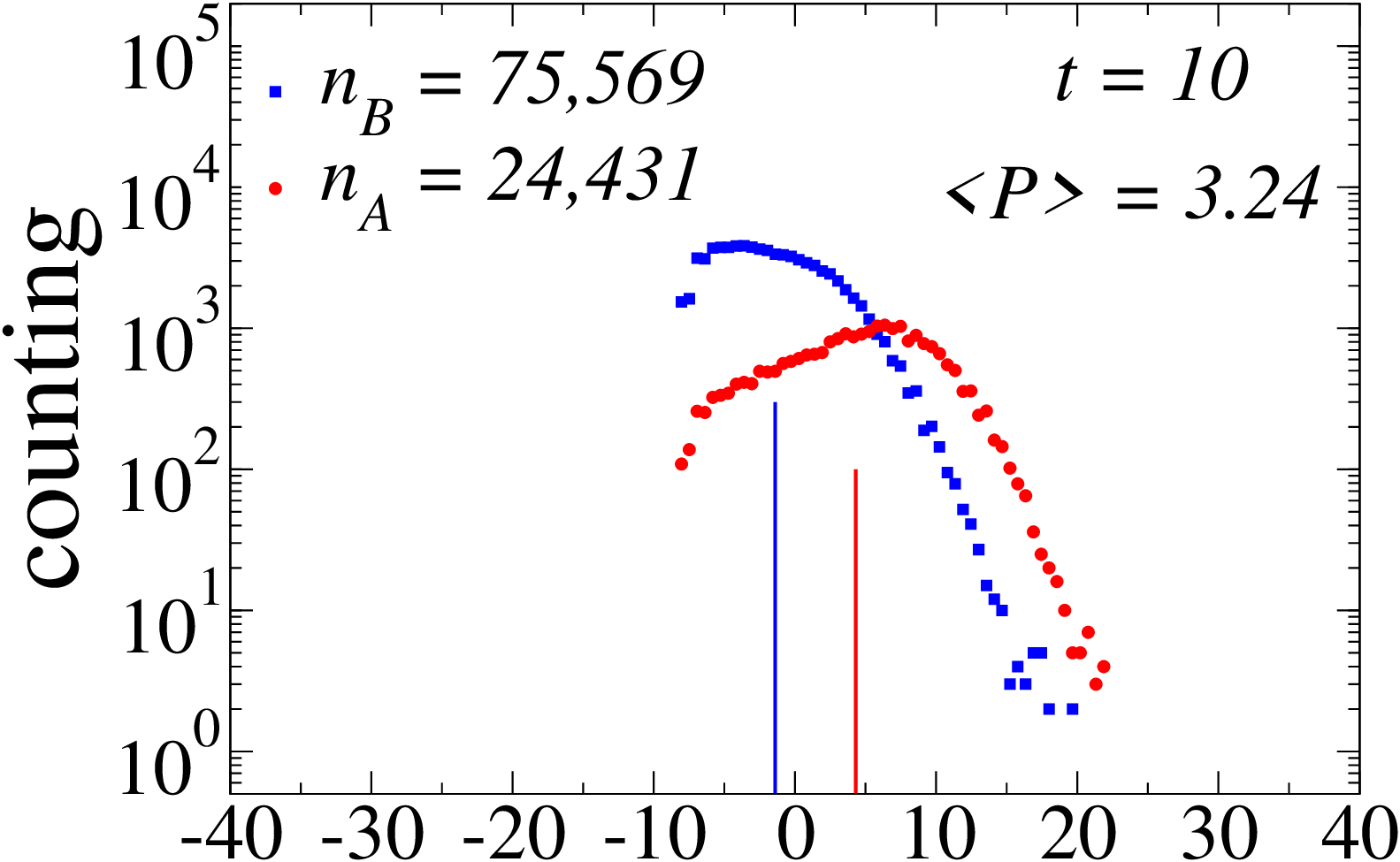}
\hspace{0.2cm}
\includegraphics[width=0.285\textwidth,angle=0]{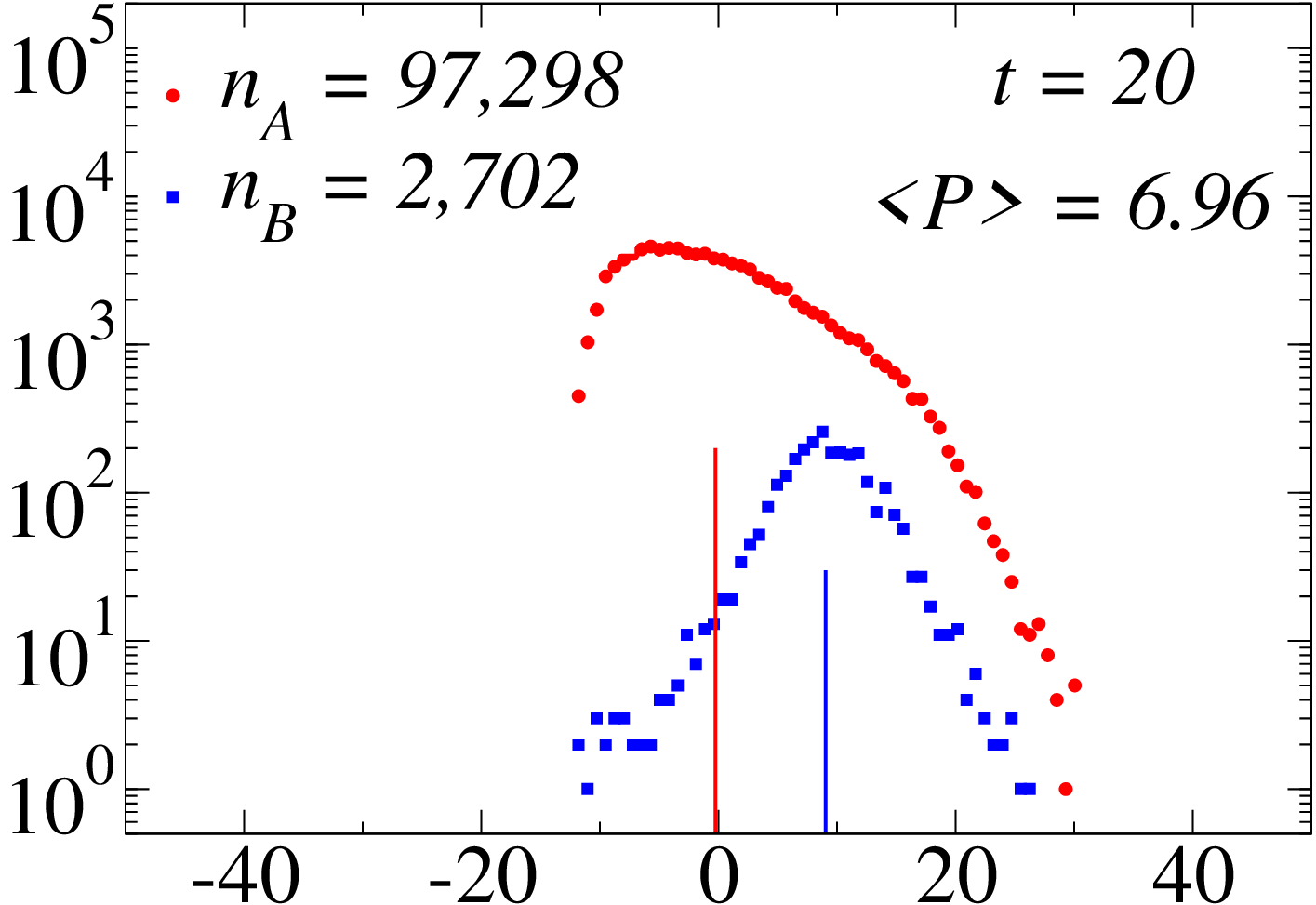}
\\
\vspace{0.48cm}
\includegraphics[width=0.3\textwidth,angle=0]{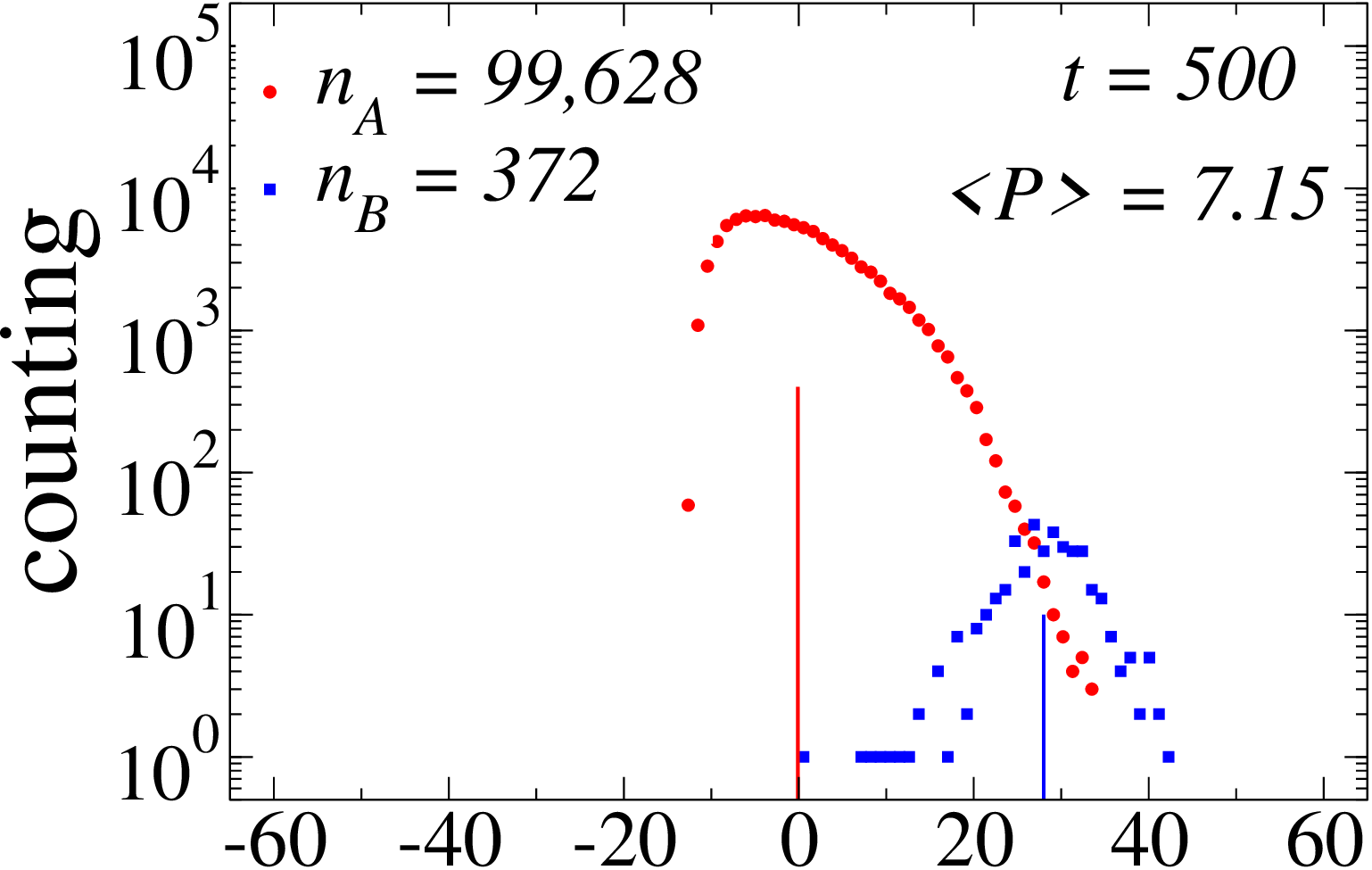}
\hspace{0.2cm}
\includegraphics[width=0.285\textwidth,angle=0]{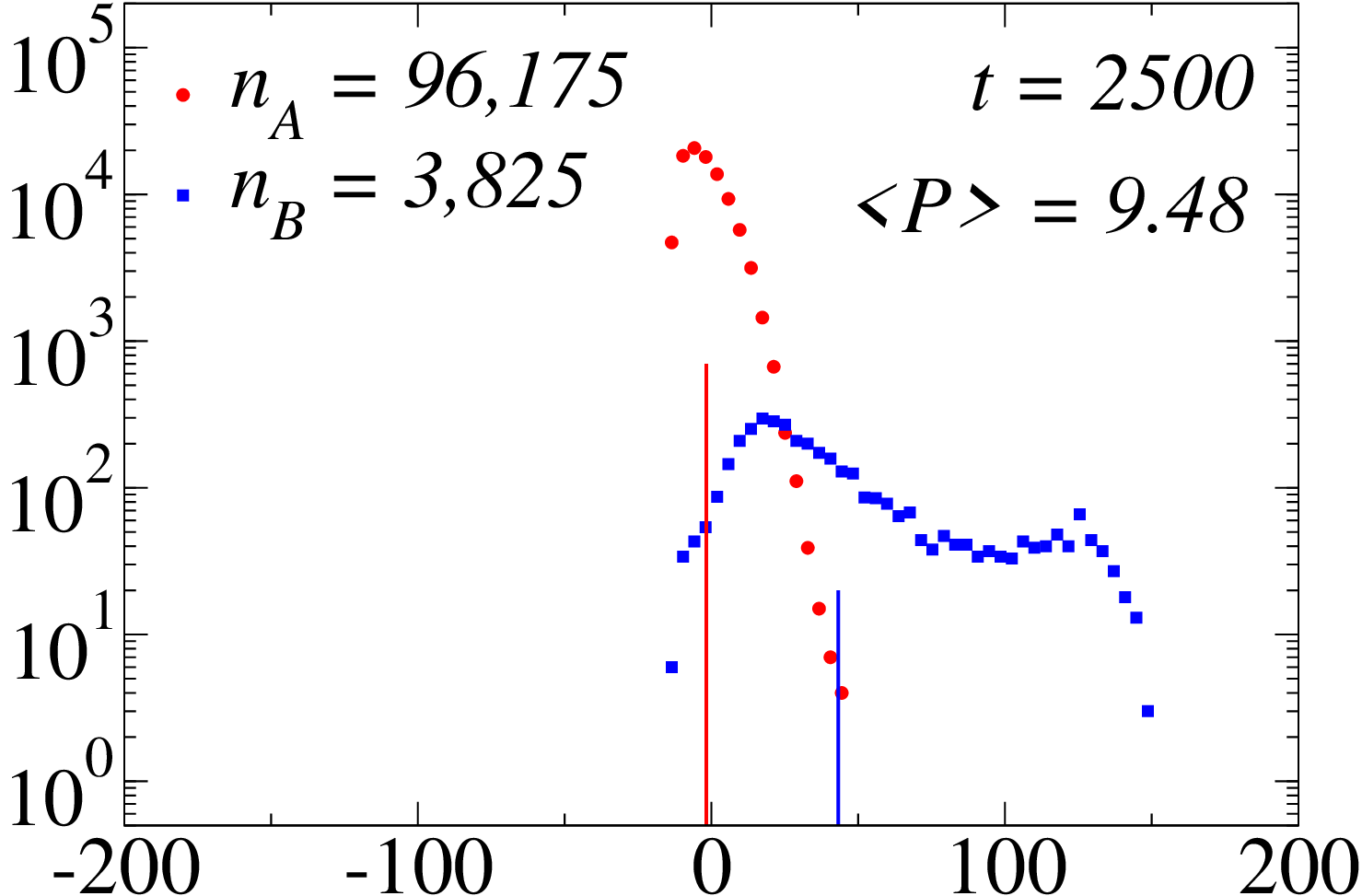}
\\
\vspace{0.52cm}
\includegraphics[width=0.3\textwidth,angle=0]{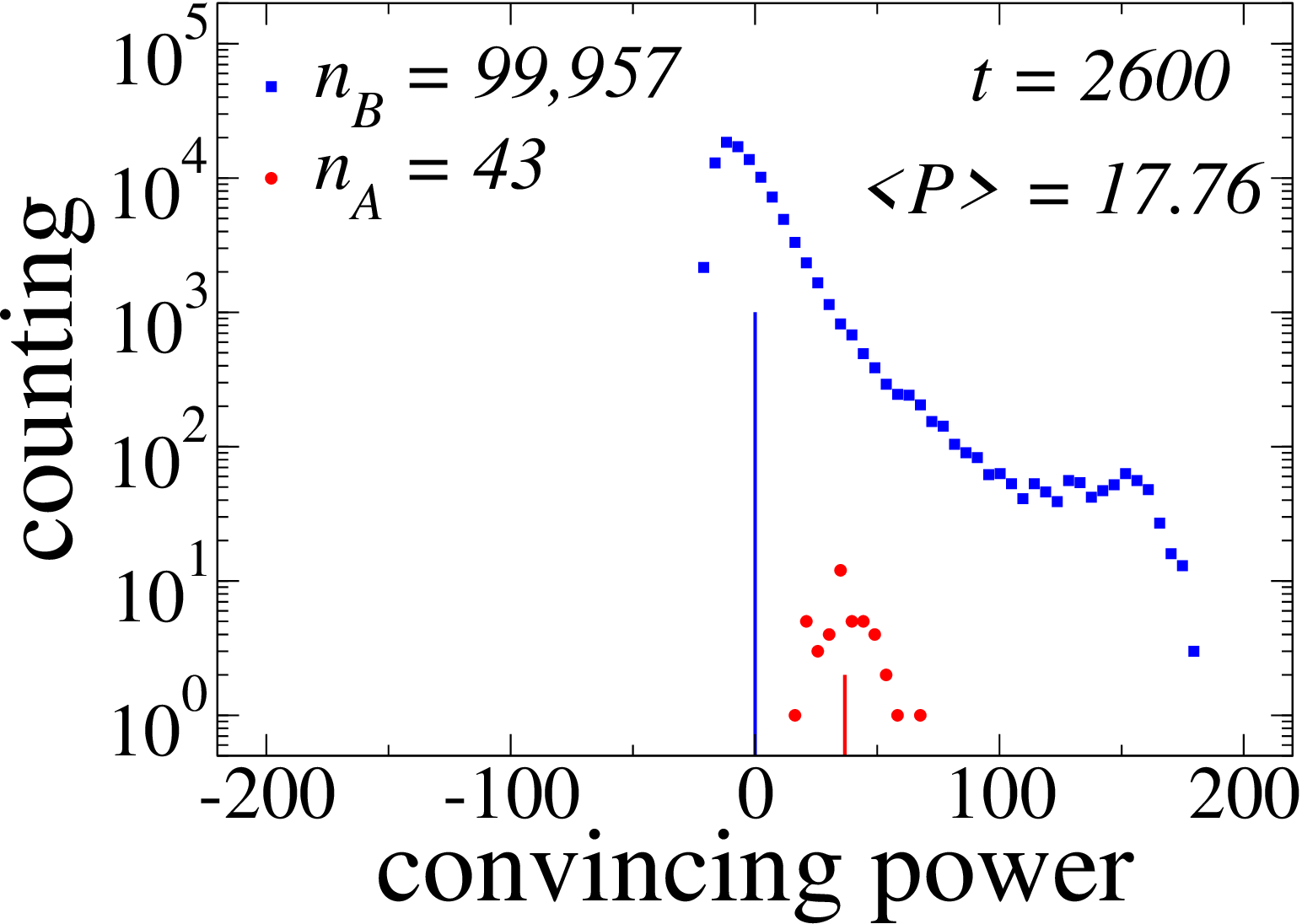}
\hspace{0.2cm}
\includegraphics[width=0.285\textwidth,angle=0]{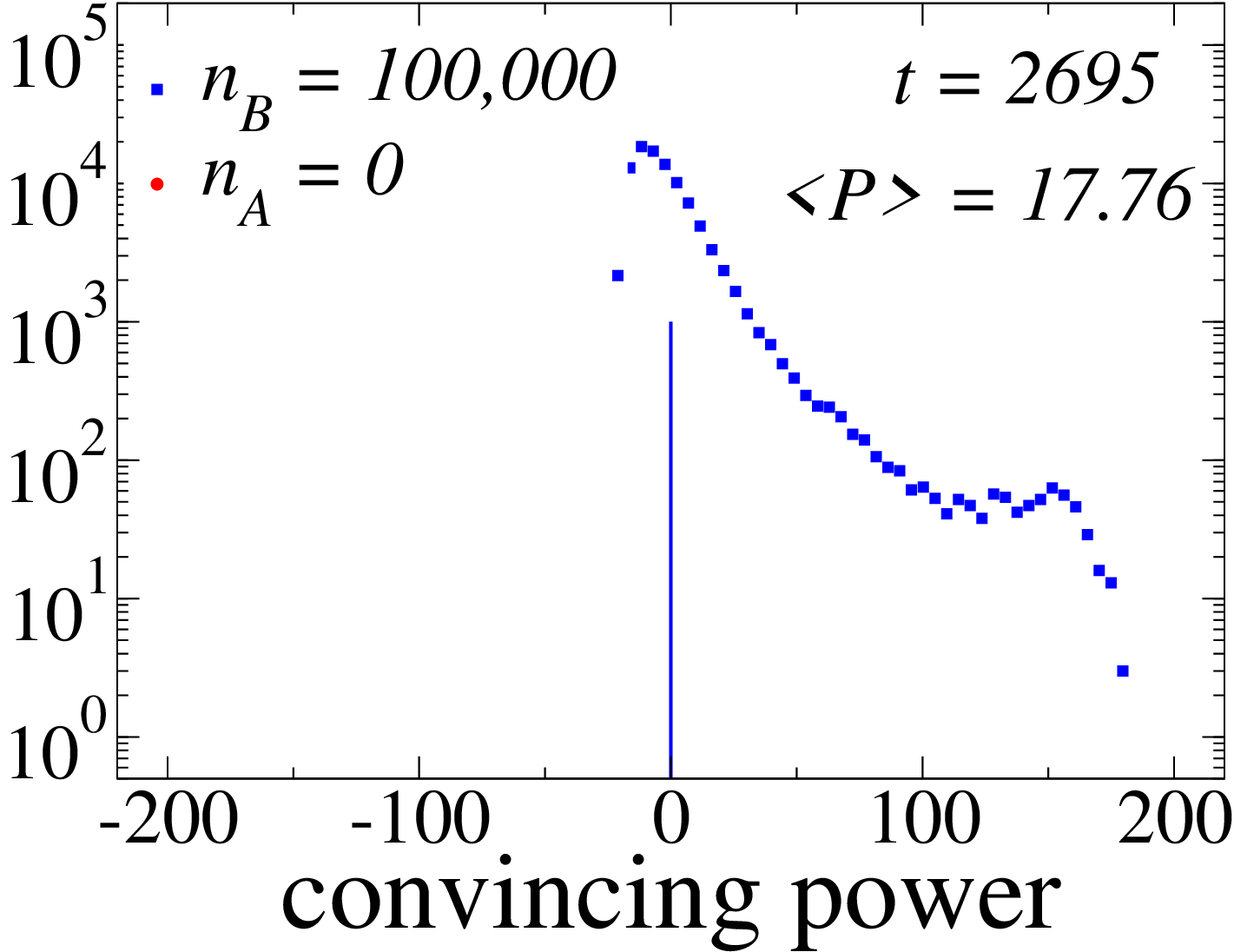}
\end{center}
\caption{(Color online) Time evolution of $A$- and $B$- distributions of convincing powers for $D=0.35$ and population size $N=10^{5}$. At beginning, the Darwinian bottleneck selection among the $A$-minority is not so strong, as can be observed by comparing the slope of the minority distribution at $t=1$ with the previous figure. The stronger the bottleneck (smaller minority), the stronger its selection pressure. Here, even under a weaker selection than in Fig. \ref{fig2}, the minority also succeeds in becoming majority between $t=10$ and $t=20$. But now the new $B$-minority is much smaller, the Darwinian bottleneck selection is now strong enough to guarantee a further majority inversion between $t=2500$ and $t=2600$. Eventually $B$-consensus is settled.}
\label{fig3}
\end{figure}

In Fig. \ref{fig3} we exhibit the time evolution of the CPs for $D=0.35$. As in the previous case, an initial minority supports $A$-opinion, but now it is not a so small minority as for $D=0.1$, Fig. \ref{fig2}. As time goes by, this minority becomes majority and later minority again, eventually becoming extinct. For $D=0.46$ (not shown), the initial majority is slightly larger than the initial minority. At beginning, the Darwinian selection bottleneck is weaker yet than in both previous cases. The consequence is the occurrence of three majority inversions. After the first inversion, the new minority becomes smaller than the initial one, but not small enough.  Only after the second inversion the minority becomes small enough to guarantee a further, definitive inversion. By gradually increasing $D$ towards $0.5$, one can observe 4 inversions, then 5, 6, etc, forming an infinite series. Similar behavior can be observed in biological models of two-species competition, in which the species with smaller population size is, under certain conditions, less susceptible to extinction than the more populous species \cite{redner2, dickman}.

\begin{figure}[t]
\begin{center}
\vspace{0.5cm}
\includegraphics[width=0.55\textwidth,angle=0]{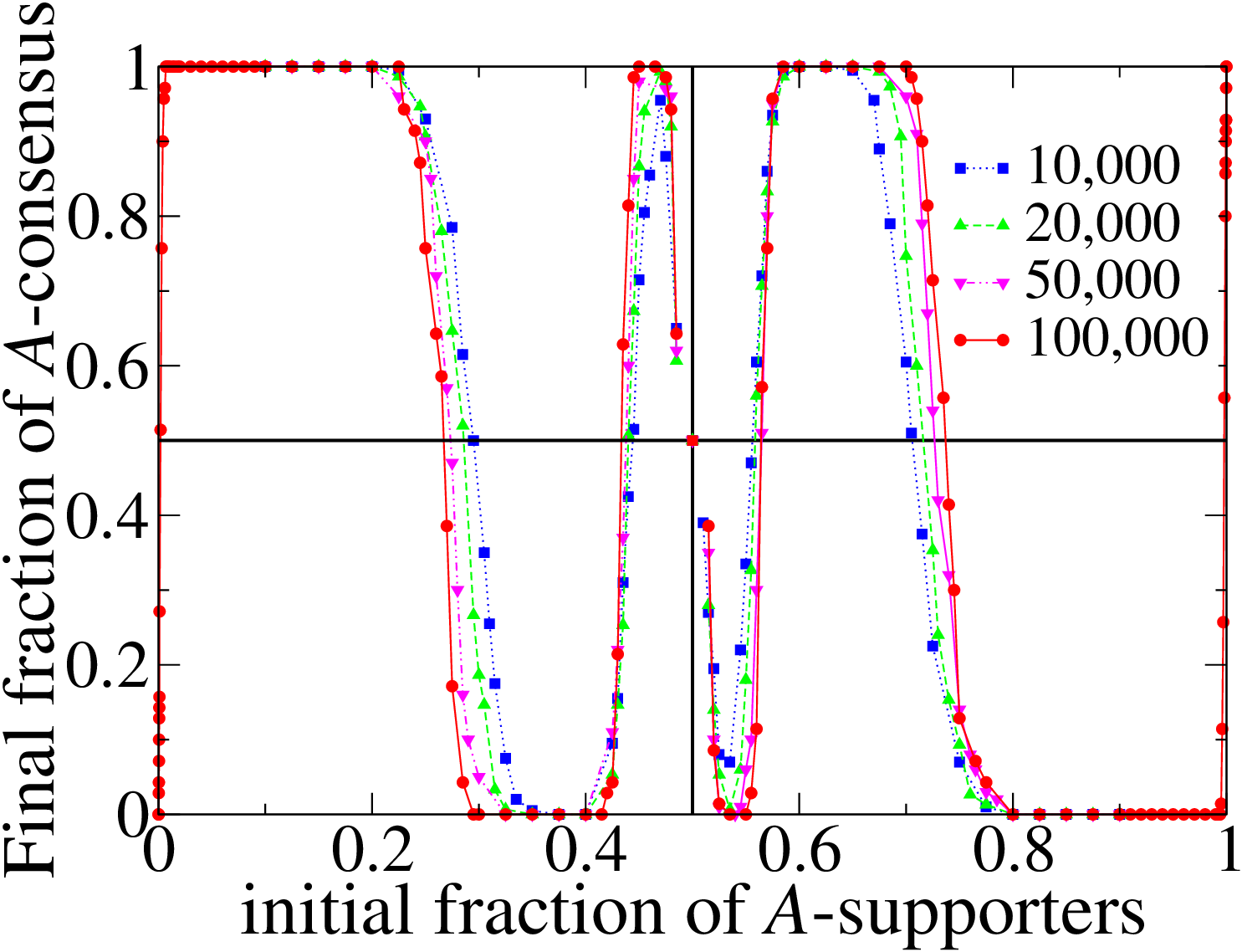}
\end{center}
\caption{(Color online) Final fraction of realizations (computer runs) which show absorbing states with $A$-consensus ($m=+1$) when the initial fraction of $A$-supporters is varied, for some population sizes $N$. Notice the symmetry with respect to the vertical and horizontal (black) lines.}
\label{fig4}
\end{figure}

To summarize the results, we plot in Fig. \ref{fig4} the final fraction $f$ of simulations which show absorbing states with $A$-consensus, when the initial fraction $D$ of $A$-supporters is varied, for some population sizes $N$. One can observe regions (plateaus) were the probability of $A$-consensus is $f=1$ and some others where this probability is $f=0$. Observe that these plateaus become ever more narrow as the limiting case $D=0.5$ is approached. Also, in the thermodynamic limit ($N \to \infty$) the smooth curves that separate the plateaus will present a straight-step, vertical behavior, i.e. the final absorbing state depends only on $D$, not on the specific initial distribution of convincing powers. Furthermore, the above-discussed majority oscillations occur more times as $D$ approaches $0.5$.


\section{Final Remarks}

To conclude, we considered a limited-persuasion mechanism in the majority-rule model \cite{galam1,galam2}. Thus, agents with a local majority opinion in a given group can be persuasive only if their average capacity of persuasion is larger than the minority's conviction. This limitation provokes the appearance of a selection rule acting in favor of the minority, a Darwinian bottleneck which opens the possibility of the initially minority opinion dominating the population in the long-time limit, provided it is sufficiently small. These findings suggest that such behaviors, typically observed in problems with biological interest like two-species competition ones \cite{redner2,dickman}, can occur also in sociological systems as the human dynamics of opinion formation.


\section*{Acknowledgments}

The authors are grateful to Ronald Dickman for fruitful suggestions, including the title. Financial support from the Brazilian funding agencies FAPERJ, CAPES and CNPq is also acknowledged.


\begin{thebibliography}{30}


\bibitem{OGlobo} Gil Cordeiro Dias Ferreira, {\it O Globo}, Rio de Janeiro, April 18, 2011.

\bibitem{galam_book}
S. Galam, \textit{Sociophysics: A Physicist's Modeling of Psycho-political Phenomena} (Springer, Berlin, 2012).

\bibitem{sen_book}
P. Sen, B. K. Chakrabarti, \textit{Sociophysics: an introduction} (Oxford University Press, Oxford, 2013).

\bibitem{loreto_rmp}
C. Castellano, S. Fortunato, V. Loreto, {\it Rev. Mod. Phys.} {\bf 81}, 591 (2009).

\bibitem{galam1}
S. Galam, {\it Physica A} {\bf 274}, 132 (1999).

\bibitem{galam2}
S. Galam, {\it Eur. Phys. J. B} {\bf 25}, 403 (2002).

\bibitem{redner}
P. L. Krapivsky, S. Redner, {\it Phys. Rev. Lett.} {\bf 90}, 238701 (2003).


\bibitem{meu_pla}
N. Crokidakis, F. L. Forgerini, {\it Phys. Lett. A} {\bf 374}, 3380 (2010).

\bibitem{redner2}
A. Gabel, B. Meerson, S. Redner, {\it Phys. Rev. E} {\bf 87}, 010101(R) (2013).

\bibitem{dickman}
R. V. dos Santos, R. Dickman, {\it J. Stat. Mech.} P07004 (2013).







\end{thebibliography}
\end{document}